\documentclass[conference]{IEEEtran}
\IEEEoverridecommandlockouts
\usepackage{cite}
\usepackage{amsmath,amssymb,amsfonts}
\usepackage{booktabs}
\usepackage{graphicx}
\usepackage{textcomp}
\usepackage{xcolor}
\usepackage{algorithm}
\usepackage{algpseudocode}
\usepackage{adjustbox}
\usepackage{multirow}
\usepackage{marvosym}

\def\BibTeX{{\rm B\kern-.05em{\sc i\kern-.025em b}\kern-.08em
    T\kern-.1667em\lower.7ex\hbox{E}\kern-.125emX}}
\begin{document}

\title{
EGNInfoLeaker: Unveiling the Risks of Public Key Reuse and User Identity Leakage in Blockchain}

\author{\IEEEauthorblockN{Chenyu Li\IEEEauthorrefmark{1}\IEEEauthorrefmark{2},
Xueping Liang\IEEEauthorrefmark{3},
Xiaorui Gong\IEEEauthorrefmark{1}\IEEEauthorrefmark{2} and Xiu Zhang\IEEEauthorrefmark{1}\IEEEauthorrefmark{2}\textsuperscript{\Letter}}

\IEEEauthorblockA{\IEEEauthorrefmark{1}Institute of Information Engineering, Chinese Academy of Sciences, Beijing, China}

\IEEEauthorblockA{\IEEEauthorrefmark{2}School of Cyber Security, University of Chinese Academy of Sciences, Beijing, China}

\IEEEauthorblockA{\IEEEauthorrefmark{3}Department of Information Systems and Business Analytics, Florida International University}

\IEEEauthorblockA{\{lichenyu1999, zhangxiu, gongxiaorui\}@iie.ac.cn, xuliang@fiu.edu}
}

\maketitle

\begin{abstract}
While Ethereum's discovery protocols (Discv4/ Discv5) incorporate robust cryptographic designs to protect user privacy, real-world deployment reveals critical vulnerabilities when users deviate from security guidelines. In this paper, we design a system called EGNInfoLeaker. Our study is the first work that uncovers widespread public key reuse across Ethereum's peer-to-peer networks - a practice that fundamentally undermines the protocol's privacy guarantees. Through systematic analysis of 300 real-world network snapshots, we identify 83 users controlling 483 service nodes via public key reuse, enabling precise de-anonymization through IP correlation. Using evidence collected by EGNInfoLeaker, our Graph-Based Identity Association Algorithm links users to network entities and generates comprehensive user profiles. For User27, it exposes the public key, IP, network ID, location (country/region/city), and ISP/ORG details. The EGNInfoLeaker system demonstrates how such cryptographic misuse transforms theoretical anonymity into practical identity leakage, exposing users to surveillance and targeted attacks. These findings establish that protocol security depends not only on sound design but also on strict user compliance. Going forward, our detection framework provides a foundation for enhancing real-world privacy preservation in decentralized networks.
\end{abstract}


\section{Introduction}
\label{sec:intro}
Ethereum (ETH)'s anonymity features play a vital role in protecting user security and privacy.
ETH is a decentralized, open-source blockchain platform that revolutionized the industry by introducing programmable smart contracts. 
ETH extends beyond simple cryptocurrency transactions (like Bitcoin) to enable developers to build and deploy decentralized applications (dApps).
By utilizing pseudonymous addresses instead of real-world identities, ETH creates a fundamental layer of protection against targeted attacks, financial surveillance, and unwanted exposure of personal assets. 
This anonymity is particularly valuable for: preventing identity-linked crypto theft, resisting transaction pattern analysis by third parties, and maintaining true financial sovereignty without reliance on traditional institutions. 
While not providing complete anonymity (as sophisticated analysis can sometimes trace activity), ETH's privacy features form a critical baseline of protection.

The ETH network operates through two distinct yet interdependent layers: the service layer and the discovery layer. 
The service layer constitutes the functional core of the network, responsible for executing smart contracts, processing transactions, and maintaining consensus through the Ethereum Virtual Machine (EVM). 
This layer ensures deterministic state transitions and enforces the protocol's business logic via globally synchronized computation. 
In parallel, the discovery layer manages the peer-to-peer (P2P) networking infrastructure, employing the Discv4 or Discv5 protocol to dynamically maintain node connectivity through a Kademlia-based distributed hash table (DHT). 
This layer facilitates node discovery, topology maintenance, and efficient data propagation while implementing cryptographic identity verification to resist Sybil, Partition, and Eclipse attacks. 
In this paper, we primarily focus on the issue of user identity leakage caused by public key reuse in the discovery layer.

Blockchain discovery layers utilize cryptographic algorithms to generate public keys as a fundamental security mechanism. 
These algorithms, such as the elliptic curve digital signature algorithm (ECDSA) with secp256k1 parameters commonly used in ETH, create mathematically-linked public/private key pairs that serve multiple security purposes. 
The public key acts as a pseudonymous network identifier while the private key remains securely stored by the node operator. 
This design provides three critical security properties: first, it enables secure node authentication through digital signatures; second, it establishes Sybil resistance by making identity creation computationally non-trivial; and third, it facilitates secure communication channels through key-derived encryption. 
The cryptographic binding between a node's identity and its key material prevents impersonation attacks while maintaining the decentralized nature of peer verification. 
However, this security model depends on proper key management practices, as the reuse of public keys across different contexts or protocols can create unintended identity linkages that compromise user privacy - the core focus of our investigation. 

Current research on blockchain discovery networks primarily investigates three interconnected aspects: (1) network attribute characterization, (2) intrinsic security vulnerabilities, and (3) service network implications. 
Studies of network attributes~\cite{Maeng2020,Maeng2021,Kim2018,Lucianna2021,Lee2020} focus on topological metrics like node distribution and stability, utilizing public keys as primary identifiers in Discv4/Discv45 protocols. 
Security analyses~\cite{Eisenbarth2022,XU2020,Marcus2018} reveal fundamental threats to the DHT structure, particularly Sybil, Partition, and Eclipse attacks that can disrupt network operations. 
Crucially, these discovery-layer vulnerabilities propagate to service networks through shared infrastructure~\cite{TaotaoWang2021Ethna,Kai2021,Zhao2024DEthna}, enabling cross-layer attacks including DoS and privacy breaches. While existing work~\cite{Caspar2021,Mikel2024,DISCNG} has examined isolated attack vectors, our study uniquely demonstrates how public key reuse creates systemic identity linkage vulnerabilities that simultaneously compromise both network layers. 
This work bridges a critical gap between protocol-level security~\cite{Sebastian2019,Hwanjo2023} and service-layer privacy concerns~\cite{Dominic2025}, revealing novel attack surfaces in decentralized network architectures.

While existing research has extensively studied protocol-level security mechanisms~\cite{XU2020,Alex2015,Irani2018} and network measurement methodologies~\cite{Mikel2024,Maeng2021,Kim2018}, a critical gap remains in understanding the security implications of public key reuse by end users - a prevalent but overlooked practice that violates cryptographic scheme guidelines. 
Current approaches predominantly rely on public keys as node identifiers due to NAT (Network Address Translation) limitations~\cite{Shang2016ChallengesII}, yet fail to address the fundamental contradiction: these same keys, designed to protect communication privacy and node holder anonymity, become vectors for identity linkage when reused across nodes. 
This user-induced vulnerability creates an unexamined attack surface where malicious actors could deanonymize node operators and track network behavior, completely compromising the privacy guarantees that public key cryptography was intended to provide in discovery protocols. 
The tension between public key reuse for operational convenience and its severe security consequences represents an understudied but crucial aspect of blockchain network security that bridges protocol design and real-world deployment challenges.

To tackle these challenges, we propose a system named EGNInfoLeaker (Ethereum Global Network Information Leaker),  which is a privacy risk detection system for blockchain networks that identifies security threats caused by public key reuse. 
It consists of four modules: a Crawler that maps network nodes, a Handshaker that verifies node services, an Integrator that correlates identities across nodes, and a Database storing collected evidence. 
By analyzing how users reuse public keys across multiple nodes, EGNInfoLeaker exposes hidden privacy vulnerabilities that undermine blockchain anonymity guarantees. 
This approach bridges the gap between cryptographic theory (where public keys should be unique) and real-world practice (where users often reuse public keys), providing new insights into identity leakage risks in decentralized networks.

Our main contributions are as follows:
\begin{itemize}
\item \textbf{New Findings.} We are the first to identify the privacy risks arising from public key reuse. 
And, this work constitutes the first systematic study of privacy leakage caused by public key reuse. 
We analyze 300 real-world network snapshots and trace 483 service nodes to 83 distinct users by detecting public key reuse, which—combined with IP correlation—enabled precise de-anonymization.
\item \textbf{New System.} We design and implement the EGNInfoLeaker system, which is a privacy-preserving detection system that identifies public key reuse in blockchain networks by crawling node data, verifying service handshakes, and correlating user identities across multiple nodes to mitigate deanonymization risks.
\item \textbf{New Algorithm.} We develop a Graph-Based Identity Association Algorithm, which is capable of clearly associating users with entities and generating detailed user profiles based on the information detected by EGNInfoLeaker. 
Taking User27 as a case study, our algorithm reveals its associated public key, IP address, network ID, country, region, city, Internet Service Provider (ISP), and organizational (ORG) information.
\end{itemize}

In Section~\ref{sec:rw}, we present related work in this field. 
In Section~\ref{sec:cc}, we introduce the core concepts and mechanisms of the Discovery network. 
Section~\ref{sec:sd} describes the design of EGNInfoLeaker. 
In Section~\ref{sec:exp}, we provide experimental evidence to support our findings. 
Section~\ref{sec:disLimit} includes a discussion of the experimental results and the limitations of EGNInfoLeaker. 
Finally, in Section~\ref{sec:conclusion}, we consolidate the principal contributions of this work.

\section{Related Work}
\label{sec:rw}
The current research on the discovery network primarily focuses on the following areas: the first is the attributes of the discovery network, the second is the security issues of the discovery network itself, and the third is how the security of the discovery network impacts the service network.

For the first category of research, studies focus on the overall graph attributes~\cite{Maeng2020,Maeng2021,Kim2018,Lucianna2021,Lee2020,Chen2020,Yue2019,Maeng2021Visualization,Zhenzhen2020,Mikel2024,DISCNG} of the discovery network, including server startup times~\cite{Maeng2020, Maeng2021, Kim2018, Lucianna2021, Lee2020, Chen2020, Yue2019, Maeng2021Visualization, Zhenzhen2020}, geographical distribution~\cite{Maeng2021, Kim2018, Lucianna2021, Lee2020, Chen2020, Yue2019, Maeng2021Visualization, Zhenzhen2020}, server stability~\cite{Maeng2020,Maeng2021,Kim2018,Lee2020}, and network message statistics~\cite{Maeng2021,Lucianna2021,Yue2019,Maeng2021Visualization}. 
These studies primarily rely on interactions with messages and other nodes in the Discv4 and Discv5 protocols to analyze the information obtained.
Typically, these studies use public keys as identifiers (IDs) to distinguish different node identities.

For the second category of research~\cite{Eisenbarth2022,XU2020, Marcus2018,Sebastian2019,Hwanjo2023,LSDAttack}, the security of the discovery network primarily depends on the security of the Discv4 and Discv5 protocols themselves, as well as the robustness of the DHT data structure used for information storage. 
The discovery network is mainly exposed to three types of Partition~\cite{Hwanjo2023}, and Eclipse attacks~\cite{XU2020,Marcus2018,Sebastian2019}. 
These attacks directly target the DHT, which can significantly degrade network efficiency, cause network partitioning, and even lead to nodes being isolated through eclipse attacks.

For the third category of research~\cite{TaotaoWang2021Ethna,Kai2021,Zhao2024DEthna,Dominic2025,Caspar2021,Mikel2024,DISCNG}, the primary focus lies in the security of the service network. 
Since the nodes of the service network are selected from those within the discovery network, the security of the discovery network directly and significantly impacts the security of the service network. 
This line of research can be broadly divided into two subcategories: one investigates attacks~\cite{Caspar2021,Mikel2024,DISCNG} that manipulate neighbor information within the DHT to compromise the service network, while the other~\cite{TaotaoWang2021Ethna,Kai2021, Zhao2024DEthna,Dominic2025,Caspar2021} examines vulnerabilities in the protocols of the service network itself that can be exploited for attacks. 
Similar to the discovery network, the service network is also susceptible to Sybil, Partition, and Eclipse attacks. 
In addition, it faces further threats such as Denial of Service (DoS) attacks and the potential leakage of sensitive neighbor relationships, raising serious privacy concerns.

This study primarily focuses on how public key reuse within the discovery network undermines its security mechanisms and the resulting negative impacts on the service network. 
Furthermore, it investigates how such public key reuse compromises both the privacy and security of the service network.

\section{Core Concept}
\label{sec:cc}
In this section, we provide an overview of the relevant background knowledge. 
First, we introduce the DHT-related protocols, Discv4 and Discv5, as well as the associated tasks and data storage structures. Then we explain how nodes perform the service handshake process and how distinct users cryptographically establish their identities.

\subsection{All about the DHT}
\begin{figure}[h]
  \centering
  \includegraphics[width=1\linewidth]{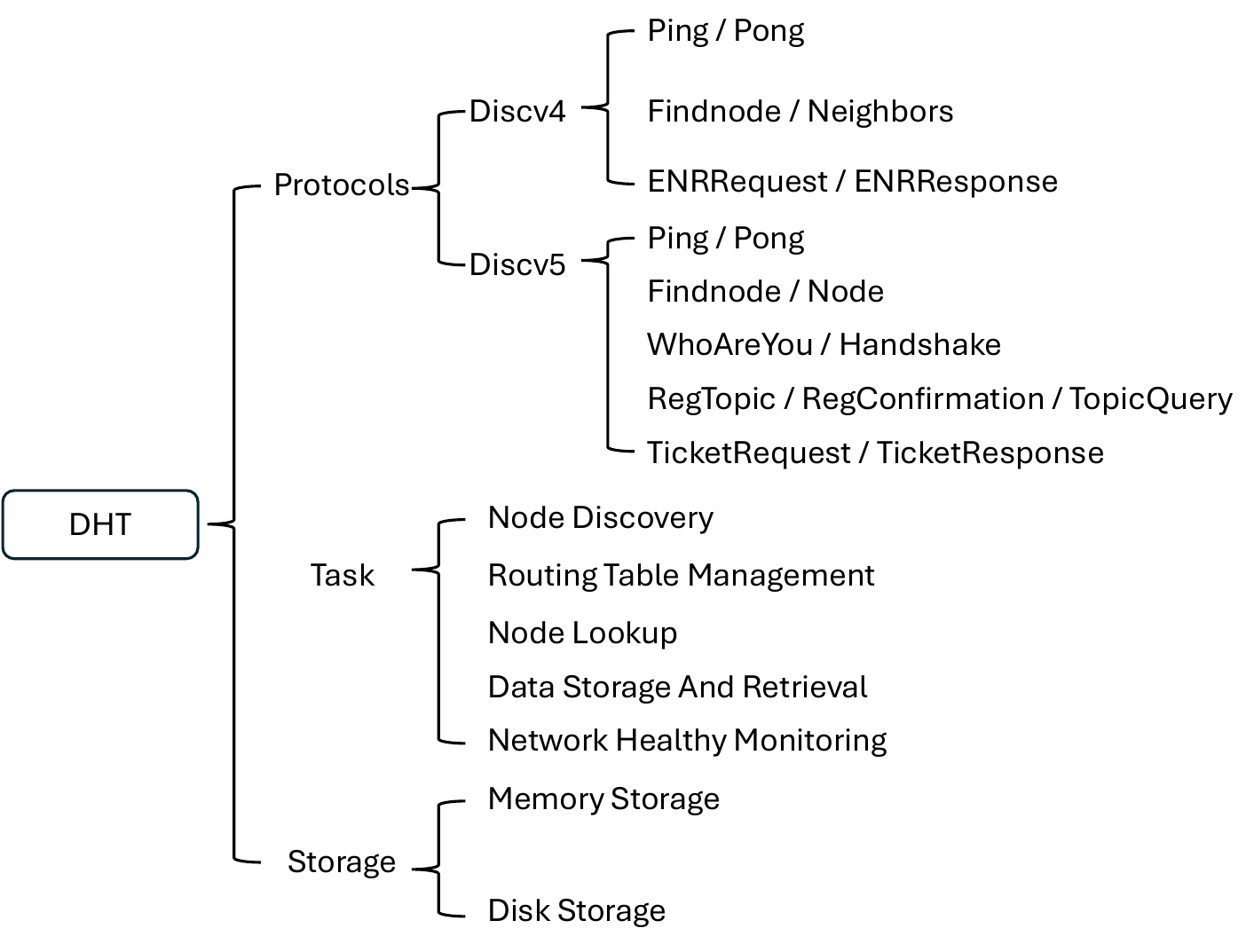}
  \caption{Protocols, tasks, and data structures related to DHT.} 
  \label{fig:concepts}
\end{figure}

A DHT is a structured key-value storage system, as shown in Figure \ref{fig:concepts}.

According to the protocol, the discovery network is categorized into the Discv4 and Discv5 networks. 
Its primary function is to guide new nodes in joining the blockchain network through the use of long-standing nodes known as bootnodes, utilizing the discovery protocol.
The Discv5 protocol serves as an upgraded version of the Discv4 protocol. 
Discv4 primarily involves three types of messages: Ping/Pong, Findnode/Neighbors and ENRRequest/ENRResponse. 
The Ping/Pong messages are used to verify the proper functioning of the Discv4 protocol on a node, while Findnode/Neighbors messages are used to obtain the Discv4 peer information of a node. 
ENRRequest/ENRResponse messages are employed to retrieve the node’s ENR.
Discv5 has five message types: Ping/Pong, Findnode/Node, WhoAreYou/Handshake, RegTopic/RegConfirmation/TopicQuery, TicketRequest/~TicketResponse.
The first three message types cover core Discv4 functions: Ping/Pong checks node liveness, FindNode/Node requests neighbors, and WhoAreYou/Handshake verifies identity. 
The last two message types, unique to Discv5, categorize services to prevent mixing, improving security and efficiency over Discv4.

The DHT implementation performs five key tasks: node discovery, routing table management, node lookup, data storage/retrieval, and network health monitoring.
Node discovery involves connecting to bootstrap nodes during startup, using Ping/Pong to verify node availability, and periodically refreshing the routing table.
Routing table management involves storing nodes both in memory and on disk, while maintaining DHT bucket positions using an LRU (Least Recently Used) strategy.
Node lookup facilitates an iterative process for service handshakes and recursive node location.
Data storage/retrieval focuses on handling lightweight metadata, such as public keys and IP addresses.
Network health monitoring periodically checks the status of nodes to ensure continuous connectivity.

In a DHT, there are primarily two storage components: memory storage and disk storage.
The memory storage component handles the run-time storage of discovery network data, including information about known nodes.
It consists of two main structures: buckets and replacement lists. 
The memory is divided into $n$ buckets (17 in the case of Go Ethereum), each with a size of $m$ (16 in Go Ethereum). 
Each bucket is associated with a corresponding FIFO (First-In-First-Out) replacement list. 
When a bucket reaches its capacity, new nodes are placed in the replacement list. 
If a node within the bucket becomes inactive, a node from the replacement list will replace it.
The disk storage component is responsible for persisting memory storage data to the hard disk, allowing the system to initialize with stored data upon restart. 
It includes three tables: the live nodes table, the failed neighbor requests table, and the seed nodes table.
The live nodes table contains nodes that have passed Discv4/Discv5 liveness checks and serves as a source for future seed nodes. 
The failed neighbor requests table records unsuccessful requests due to network or data issues, supporting memory storage maintenance. 
The seed nodes table holds protocol-compliant nodes used to initiate new discovery requests within the memory storage.

In brief, both the Discv4 network and the Discv5 network have their own separate instances of DHT programs, and the data from Discv4 and Discv5 are not mixed within the same DHT.
The data stored in the DHT includes the Ethereum Node Record (ENR) and related information. The key insight is that: 
\textbf{A node will only exist once within the entire DHT, and the node’s public key will serve as its unique identifier within the DHT.} 
In other words, regardless of the service or different services, only one node’s information associated with a given public key will be stored in the DHT.

\subsection{Service Handshake}
In order to establish a service connection with nodes of the same service, a node iteratively retrieves node information from the DHT and attempts to perform a service handshake. 
Currently, the node information stored in the DHT does not include service-related information, so the node directly proceeds with the service handshake using the retrieved node information.

In the Discv4 network, the services exhibit a high degree of similarity~\cite{Kim2018,Zhenzhen2020,Lin2021,Maeng2021Visualization}, as most of the services in Discv4 are based on code from Go Ethereum, such as those used by Ethereum Execution, BSC, Polygon, and Fantom.
The entire service handshake process can be roughly divided into three steps: (1) Establish a TCP connection using the node information from the DHT. 
(2) Establish an encrypted communication connection using the Recursive Length Prefix extension (RLPx) Transport Protocol over the TCP connection. 
(3) In RLPx, the two nodes exchange their service information tuples (Protocol Version, Network ID, Genesis Block Hash, Fork ID); if they are identical, the service handshake is completed.
Any discrepancy in even a single element of the tuple will result in the services being classified as different.

Compared to Discv4, the services in Discv5 are also diverse, with the handshake process being even more varied~\cite{Abhishek2023,Mikel2024,Kim2018,Lin2021}. 
In Discv5, the services are primarily driven by Ethereum Consensus, with a small portion from Ethereum Execution, as well as other services utilizing Discv5. 
Unlike Discv4, different service handshakes in Discv5 do not follow a similar process. 
This is partly due to the diversity of configuration options in the libp2p network library protocol stack, and partly because of the variety of service protocol handshakes in Discv5. 
Only after the network protocols are fully aligned can the handshake proceed, allowing both parties to exchange their service information.

Furthermore, even between two identical services, there may still be cases where the handshake fails. 
To ensure the security and stability of the blockchain network, most services check their network connection limits before completing the service handshake when establishing a TCP connection. 
These limits restrict the total number of service connections a node can establish, and these connections are further categorized into inbound and outbound connections. 
When a new connection request exceeds the connection limits, the handshake will fail directly.

\subsection{User Identity}
\begin{figure}[h]
  \centering
  \includegraphics[width=1\linewidth]{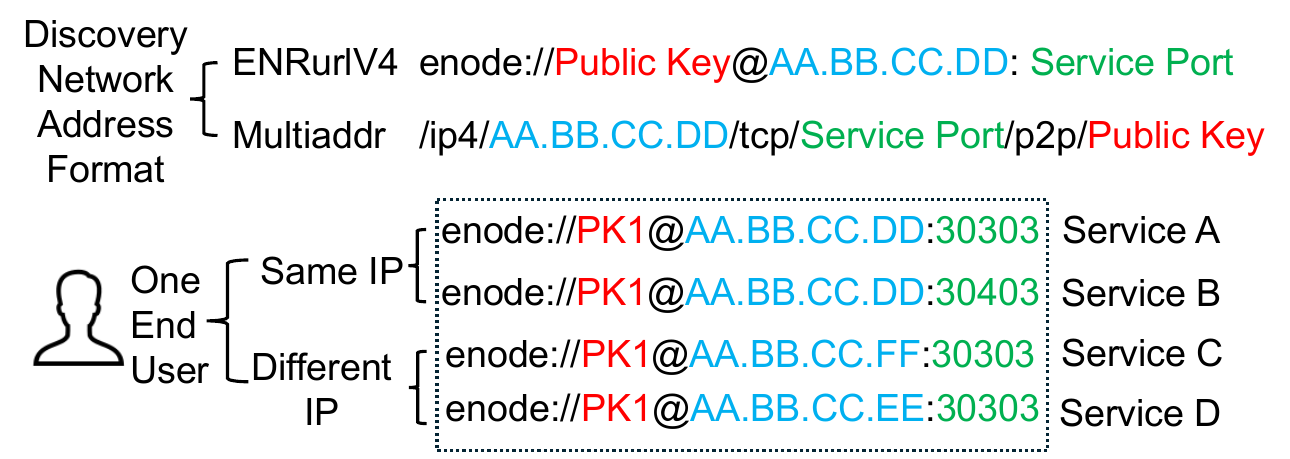}
  \caption{The top part shown the discovery network address format. When an end user uses a public reuse across multiple nodes, the DHT’s use of the public key as a unique identifier causes these nodes to be indistinguishable in the DHT, ultimately resulting in a single ENR.}
  \label{fig:concept2} 
\end{figure}

A node’s address within the Discovery network is derived from its ENR, as shown in the top part of Figure \ref{fig:concept2}.

In order to establish their identities within the discovery network, nodes generate a Secp256k1 public/private key pair. 
The public key is used to uniquely identify the node within the discovery network, while the private key is employed to establish encrypted connections when attempting to initiate service connections with other nodes.

The ENR encapsulates essential information about the node, including its Secp256k1 public key, IP address, discovery protocol port, and service port.
Information about neighboring nodes is stored in a DHT. 
In Discv4, the ENR is transformed into an address format known as ENR URL version 4 (\emph{ENRurlV4}), whereas in Discv5, the ENR is converted into an address format called \emph{Multiaddr}. 
These two formats represent different address representations of the ENR.

\section{Impact of Public Key Reuse}
\begin{figure}[h]
  \centering
  \includegraphics[width=1\linewidth]{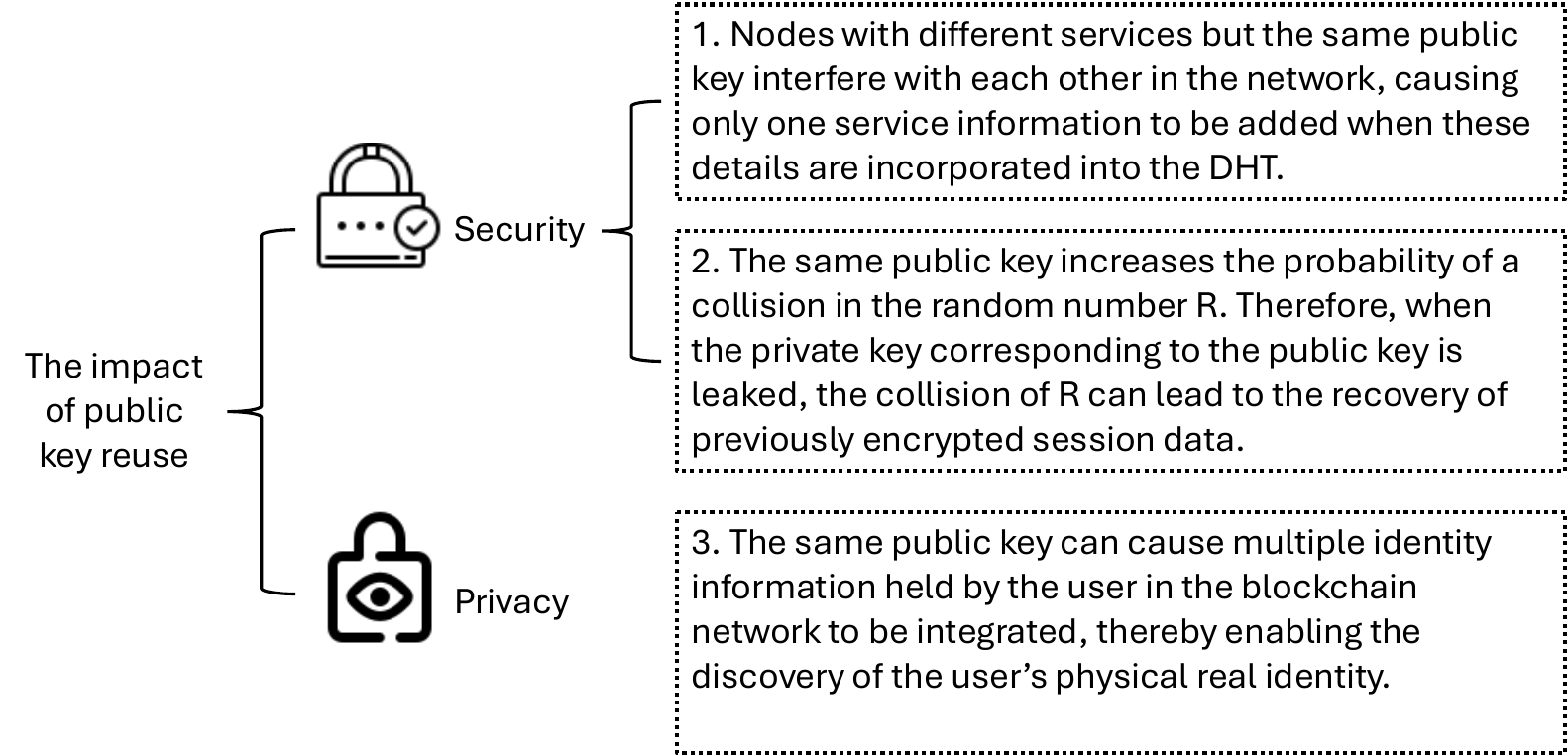}
  \caption{Public key reuse has two security impacts and one privacy impact.} 
  \label{fig:ip} 
\end{figure}

In this section, we introduce the processes in which the public key is involved. Then we analyze the potential threats to the network’s security and privacy posed by public key reuse in these processes, as shown in Figure \ref{fig:ip}. 

This public key serves two main purposes. 
The first is to act as the unique identifier of a node within the network, and the second is to participate in the encryption of the communication channel between two nodes through RLPx.
For the first purpose, the public key functions as the unique identifier used to verify a node’s identity within the DHT, preventing nodes from appearing multiple times in the DHT. 
At the same time, the public key is explicitly encoded into the ENRurlV4 and Multiaddr, as shown in the top part of Figure~\ref{fig:concept2}, allowing other nodes to quickly utilize the public key when connecting to this node and facilitating the broadcasting of information about this node to other nodes.
For the second purpose, the public key serves as a long-term key in the Elliptic Curve Integrated Encryption Scheme (ECIES) process when establishing an RLPx encrypted channel with other nodes. 
After ECIES, a temporary key is generated to encrypt the session content.

\textbf{Security.} Public key reuse can have two significant impacts on the security of nodes within the network. 
(1) The first impact is that interference among multiple services under the same user can hinder sufficient inbound connections for most services. 
According to DHT rules, using the same public key across different services (A, B, C) leads to conflicts in the Discv4 and Discv5 addresses within the DHT. 
When service A’s information reaches a DHT node, the addresses for services B and C are excluded from the node’s DHT. 
Although the same user’s services may be visible to a node, addresses tied to the same node are not added to the DHT. For instance, the DHT for service A will not include information from services B or C. 
Inbound connections typically make up two-thirds of total connections, so this situation can disrupt node synchronization. 
A lack of node connections increases susceptibility to Sybil, Partition, and Eclipse attacks.
(2) The second security impact is that public key reuse may compromise the confidentiality of historical node communications. 
In the ECIES scheme, a random number $R$ is selected for each session and discarded after use, serving as a parameter for subsequent encryption. 
Under normal circumstances, if a private key is leaked, historical encrypted communications remain secure since $R$ is not stored. 
However, if $R$ is reused, encrypted content from prior sessions using the same $R$ can be decrypted. 
Public key reuse increases the likelihood of a collision in $R$, reducing the time complexity from $2^n$ to $2^{n/2}$ due to the birthday attack threat.

\textbf{Privacy.} Public key reuse can severely undermine user privacy in the blockchain network. 
The internet operates on a TCP/IP framework, where nodes typically reveal only that a certain IP address is running a service using Discv4 or Discv5, requiring additional handshakes for service identification. 
However, with the limited availability of IPv4 addresses and the widespread use of NAT, IP addresses alone are insufficient for user identification. 
Consequently, most studies use public keys to distinguish nodes. 
When multiple public keys share the same IP, it is difficult to ascertain if they belong to the same user. 
Reusing public keys, which are cryptographically unique, enables the correlation of distinct identities, allowing identification of services, associated IP addresses, and, by observing node behaviors, the construction of a user profile. As shown in the bottom part of Figure \ref{fig:concept2}, \textbf{when an end user uses a public reuse across multiple nodes, the DHT’s use of the public key as a unique identifier causes these nodes to be indistinguishable in the DHT, ultimately resulting in a single ENR.}
This leads to the exposure of which services a user operates and which servers they control, compromising privacy by linking multiple identities to a single user.

In this paper, we first validate the security impact of public key reuse across multiple services (\ref{RQ1}), then quantitatively assess its effects on user privacy (\ref{RQ2}, \ref{RQ3}).

\section{System Design}
\label{sec:sd}
\begin{figure}[h]
  \centering
  \includegraphics[width=1\linewidth]{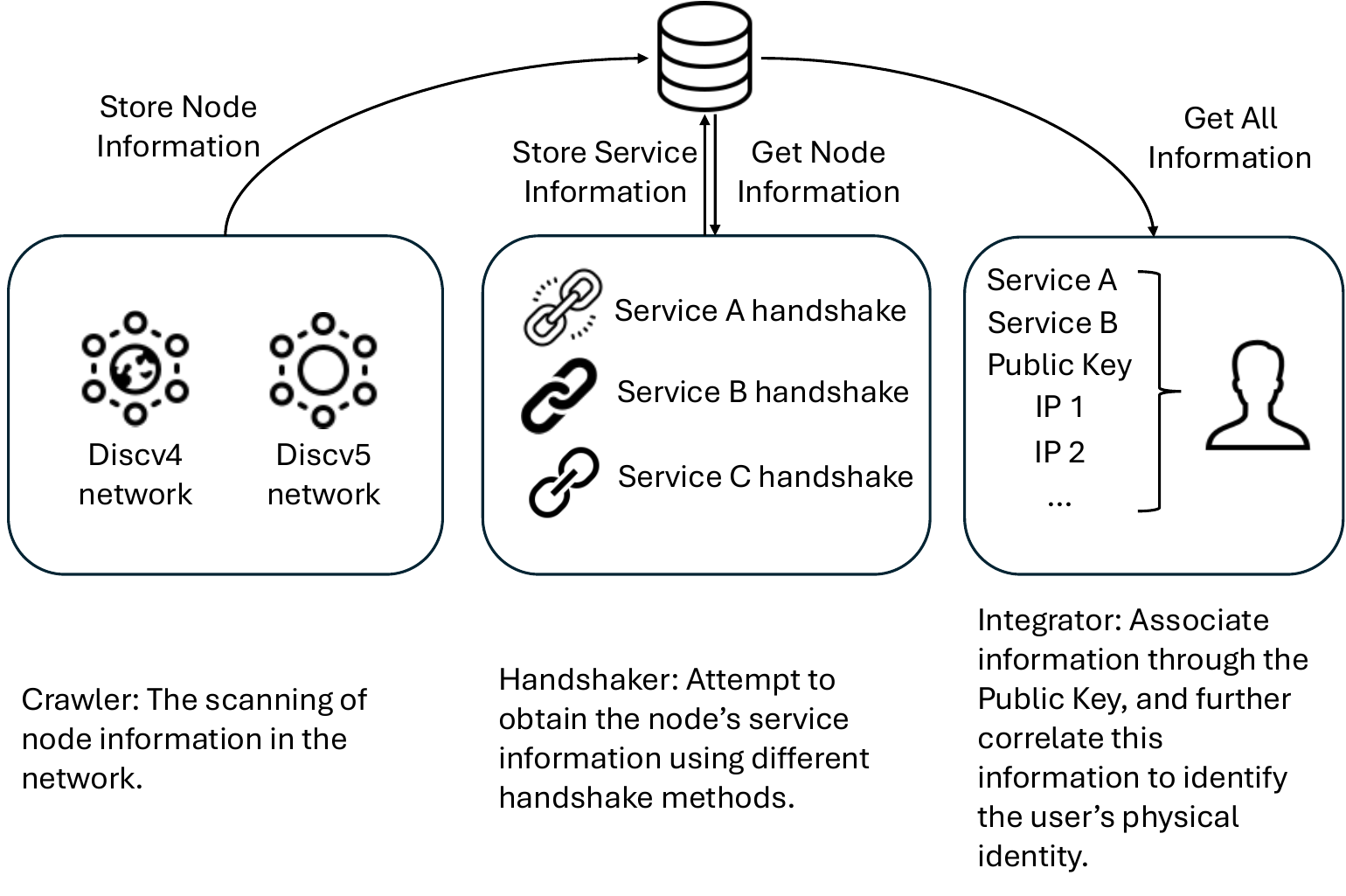}
  \caption{The overall design diagram of EGNInfoLeaker.}
  \label{fig:ov} 
\end{figure}

In order to detect misuse phenomena within the blockchain network and ultimately safeguard user privacy, we propose the EGNInfoLeaker system, as shown in Figure~\ref{fig:ov}.
This system consists of four modules: Crawler, Handshaker, Integrator, and Database. The Crawler is primarily responsible for crawling the discovery network and generating network snapshots. 
The Handshaker attempts to perform service handshakes with the nodes in the snapshot to probe the actual services provided by these nodes. 
The Integrator attempts to integrate user identities and correlate all services and node  behaviors associated with these users. 
The Database is responsible for storing the data used by the previous three modules.

\subsection{Crawler}
The Crawler is primarily responsible for periodically crawling the entire Discv4 and Discv5 networks and generating snapshots. Our Crawler uses similar techniques to most related research~\cite{Maeng2020,Maeng2021,Kim2018} in crawling networks. 

For the Discv4 network, we mainly utilize the FindNode Packet Message to send requests to target nodes.
The target node, based on the “target” parameter in the message, will return information about the 16 closest nodes. 
Each node’s DHT can accommodate up to 272 nodes, so the probability of obtaining different nodes with a single FindNode Packet Message is $16/272$. 
If we send the FindNode Packet Message to the node n times, the number of distinct nodes obtained is denoted as $\mathbb{E}$. 
The formula for $\mathbb{E}$ is $\mathbb{E} = 272 \left( 1 - \left(\frac{256}{272}\right)^n \right)$.
Therefore, when considering the number of network packets and the expected outcome, sending 40 FindNode Packet Messages to a node is expected to yield approximately 92\% of its neighboring information.

For the Discv5 network, we similarly use the FindNode Packet Message to send requests to the target node. 
The target node will return the 16 nodes that are closest to the distance specified in the FindNode Packet Message. 
According to the design of the DHT, each bucket contains nodes at different distances. 
Therefore, sending 17 FindNode Packet Messages is expected to yield all the peer information of the node.

\subsection{Handshaker}
Handshaker is primarily responsible for performing service handshakes with nodes to determine the services they are running. 
While the Crawler can obtain a global snapshot of the Discv4 and Discv5 networks over a period of time, this snapshot does not contain any service-related information. 
Therefore, Handshaker attempts to perform handshakes with all the nodes in the snapshot to retrieve their service information. 
Handshaker implements most of the service handshake methods used in related measurement studies~\cite{Miller2015DiscoveringB,Maeng2021,Yitao2020,Kim2018,Lucianna2021,Lee2020,Chen2020,Yue2019,Maeng2021Visualization}. 
Since the handshake code in the Discv4 and Discv5 networks’ services typically does not change, we are able to obtain service information for the majority of nodes in these networks. 
For other minor services existing in the Discv4 and Discv5 networks, as with other measurement tasks~\cite{Maeng2021,Lucianna2021,Maeng2021Visualization}, we only record node information without attempting to perform handshakes.

\subsection{Integrator}
The Integrator is primarily responsible for analyzing the information gathered by the Crawler and Handshaker, and for using our proposed Graph-based Identity Association Algorithm to identify users involved in public key reuse. 
The Integrator employs various public databases, such as GeoIP~\cite{geoip_2025} and Reverse Lookup~\cite{ReverseDNS}, to enhance the discovery of user-associated identity information. 
This enables the complete reconstruction of a user’s physical network identity and blockchain network identity, thereby compromising blockchain privacy.

The Graph-based Identity Association Algorithm, as shown in Algorithm~\ref{alg:gbiaa}, utilizes data from the Crawler, Handshaker, and external sources as input parameters. 
The algorithm then returns a graph representing this information, where the attributes are represented as nodes, and the information associations within the network are represented as edges. 
\textbf{If, in the resulting graph, there are weakly connected components that are significantly larger than expected, these components correspond to users who are misusing public keys.}
Furthermore, these weakly connected components directly indicate that the privacy of these users has been severely compromised.
\begin{algorithm}[h]
\caption{Graph-Based Identity Association Algorithm}
\label{alg:gbiaa}
\begin{algorithmic}[1]
\State \textbf{Input:} node\_list, node\_service\_dict, outer\_sources\_list
\State \textbf{Output:} ret\_graph
\State Initialize ret\_graph as an empty graph
\For{each node in node\_list}
    \If{node not in ret\_graph}
        \State ret\_graph.add\_node(node)
    \EndIf
    \If{node\_service\_dict[node] is not None}
        \State service $\gets$ node\_service\_dict[node]
        \If{service not in ret\_graph}
            \State ret\_graph.add\_node(service)
        \EndIf
        \State ret\_graph.add\_edge(node, service)
    \EndIf
    \For{each outer\_server in outer\_sources\_list}
        \State node\_infos $\gets$ outer\_server.query(node)
        \For{each node\_info in node\_infos}
            \If{node\_info not in ret\_graph}
                \State ret\_graph.add\_node(node\_info)
            \EndIf
            \State ret\_graph.add\_edge(node, node\_info)
        \EndFor
    \EndFor
\EndFor
\State \textbf{Return} ret\_graph
\end{algorithmic}
\end{algorithm}

\subsection{Database}
The task of the Database is to store the intermediate data from the first three modules as well as the final results. 
The Database stores snapshots of the Discv4 and Discv5 networks, the address information of nodes within the network, the service information of nodes after successful handshakes, and ultimately, this data will be used as input for the Integrator. 
Finally, the Database stores the user profiles with compromised privacy generated by the Integrator.

In a word, EGNInfoLeaker is a privacy-preserving detection system that identifies public key reuse in blockchain networks by crawling node data, verifying service handshakes, and correlating user identities across multiple nodes to mitigate deanonymization risks.

\section{Experiments}
\label{sec:exp}
In this section, we conduct three experiments to answer the following three research questions.
\begin{itemize}
\item \textbf{RQ1:} Do multiple different services using the same public key actually have a significant impact on the network?
\item \textbf{RQ2:} What are the perceived results of the EGNInfoLeaker’s Crawler and Handshaker modules in the network, and what information do these results reveal about the network?
\item \textbf{RQ3:} What public key reuse instances does the EGNInfoLeaker’s Integrator identify, and which blockchain and physical device information does it associate and recover for the users?
\end{itemize}

\noindent{\bf Ethical Considerations.}
For the first experiment, the entire experiment is conducted within a private Docker network that we set up, completely isolated from real-world blockchain networks.  
For the second experiment, the number of network messages sent by the EGNInfoLeaker’s Crawler and Handshaker modules within a unit of time is less than 1\% of the network communication messages required for normal node communication, ensuring that it does not have a significant impact on the real-world physical network.  
For the third experiment, we apply hash salting protection to the user information discovered and de-anonymized by the EGNInfoLeaker’s Integrator, and we guarantee that no user personal information is disclosed in the paper.

\subsection{Validation Experiment of Public Key Reuse}
\label{RQ1}
The validation experiment consists of two sub-experiments. 
Sub-experiment 1 (Exp. 1) demonstrates in a fully isolated private blockchain network that only the information of one node, among different services using public key reuse, is added to the node’s DHT. 
Sub-experiment 2 (Exp. 2) shows that after public key reuse, nodes entering the network face significant difficulty in establishing inbound connections.

\noindent{\bf Experiment Setup.} Our platform is a 64-core CPU, 128 GB RAM, 4 TB NVMe SSD machine with Ubuntu 24.04.
We use Kubernetes to manage the private network, allocating a maximum of two nodes per subnet (C-class IP). 
The main stream clients currently include Besu, Reth, Teku, Geth, Nethermind, and Erigon. 
Their implementations of the DHT are relatively consistent, so the specific client implementation used in this simulation environment does not affect the final experimental results.
We chose Geth@1.15 as the client version for the experimental nodes.

\subsubsection{Exp. 1}
This experiment evaluates how reusing identical public keys across nodes influences the performance of discovery protocols (Discv4 and Discv5) and the resulting DHT structure, focusing on node connectivity and data dissemination.
Three network scales (10, 100, and 300 nodes) are tested, each under two protocol settings: Discv4-only and Discv5-only.
All nodes share the same public key but differ in service information (e.g., IP and port).
The primary evaluation metric is the size of the largest weakly connected component (WCC) in the generated DHT graph.
Results are shown in Table~\ref{tab:valexp1}.

\begin{table}[h]
\centering
\caption{The weakly connected components with the maximum number of nodes in the final discovery network.}
\renewcommand{\arraystretch}{1}
\begin{tabular}{cccccc}
\toprule
Protocol &  Nodes & Services & 
\begin{tabular}[c]{@{}c@{}}Public\\ Key \end{tabular} &
\begin{tabular}[c]{@{}c@{}}Weakly\\ Connected\\ Components \end{tabular} & \begin{tabular}[c]{@{}c@{}}Max\\ Weakly\\ Connected\\ Components \end{tabular}\\ 
\hline
\multirow{3}{*}{Discv4} & 10 & 10 & 1 & 10 & 1 \\
 & 100 &  100 & 1 &  100 & 1 \\
 & 300 & 300 & 1 & 300 & 1 \\ \hline
\multirow{3}{*}{Discv5} & 10 & 10 & 1 & 10 & 1 \\
 & 100 & 100 &1   & 100 & 1 \\
 & 300 & 300 &1 & 300 & 1 \\
\bottomrule
\end{tabular}
\label{tab:valexp1}
\end{table}

The experimental results show that all nodes fail to add their own service records into their respective DHTs.
This occurs because the DHT implementation uses the public key as the sole indexing attribute.
Since all nodes share the same public key, the DHT cannot distinguish between them, leading to failures in service registration.
Consequently, the networks form weakly connected components that match the number of nodes, but the DHTs themselves remain empty.
Furthermore, no discernible difference exists between the performance of Discv4 and Discv5 under these conditions.
These results demonstrate that reusing a single public key across multiple nodes disrupts the intended behavior of the DHT and prevents proper network formation and data storage.

\subsubsection{Exp. 2}
This sub-experiment examines whether two nodes operated by the same user, each with a distinct IP address but sharing an identical public key, can simultaneously establish and sustain their presence in a DHT. The primary focus is on the second node's (Node~2) ability to disseminate its presence and receive inbound connections under this key-sharing configuration.

A total of 200 nodes are deployed, with 100 each for Service~A and Service~B. 
To avoid DHT congestion beyond the known limit of 272 nodes, the node count is capped. 
One user controls two additional observed nodes—Node~1 (Service~A, IP~A) and Node~2 (Service~B, IP~B)—both initialized with the same public key, \texttt{KEY}. 
All nodes are launched, but only Node~1 is activated initially. 
After one hour, Node~2 starts. Connection metrics (inbound/outbound) for both nodes are recorded 30 minutes later.
Each node's maximum peer count is limited to 50 (16 outbound, 34 inbound), reflecting realistic Ethereum Mainnet configurations. The chosen network size and experiment duration (90 minutes) ensure stability and complete peer discovery, as 10 minutes typically suffice for convergence at this scale.

The experimental design anticipates that Node~1, as the first to join, will successfully integrate into the DHT and establish stable connections. However, due to the shared public key configuration, Node~2 is expected to be misidentified as Node~1, which would prevent it from effectively broadcasting its distinct presence and establishing inbound connections.

\begin{table}[h]
\centering
\caption{The number of inbound and outbound connections of node 1 and node 2 in the end. The abbreviation “Sv.” stands for “Service”.}
\renewcommand{\arraystretch}{1}
\begin{tabular}{cccc}
\toprule
Protocol & 
Observed Node & 
\begin{tabular}[c]{@{}c@{}} Inbound\\ Connections\end{tabular} & \begin{tabular}[c]{@{}c@{}}Outbound\\ Connections \end{tabular}\\ \hline

\multirow{2}{*}{Discv4}  & Node 1 (Sv.A, KEY@IP A) &  34 & 16 \\
                         & Node 2 (Sv.B, KEY@IP B)&   0 & 16 \\ \hline
\multirow{2}{*}{Discv5}  & Node 1 (Sv.A, KEY@IP A) &  34 & 16 \\
                         & Node 2 (Sv.B, KEY@IP B) &   0 & 16 \\
\bottomrule
\end{tabular}
\label{tab:valexp2}
\end{table}

The results in Table~\ref{tab:valexp2} validate the hypothesis.
Node 1 reliably forms 34 inbound and 16 outbound connections, demonstrating full DHT integration.
In contrast, Node 2 lacks inbound connections despite outbound reachability, indicating its identity is unrecognized by other Service B nodes.
This stems from public key duplication: Node 2’s announcements are conflated with Node 1’s existing DHT entry, causing its address data to be ignored or overwritten.
The experiment reveals a key vulnerability in DHT-based networks—public key reuse can cause node invisibility or partitioning due to key collisions, emphasizing the necessity of unique keys per node.

\subsection{Measurement Results in Real-world Network}
\label{RQ2}
The measurement experiment mainly uses a Crawler to measure the Discv4 and Discv5 networks and attempts to handshake with these nodes as much as possible using the Handshaker. 
The network measurement by the Crawler started at 10:14 UTC on March 29, 2025, and a network snapshot is generated approximately every 30 minutes. 
We have accumulated and analyzed the network data from 300 consecutive snapshots.

\begin{figure}[h]
  \centering
  \includegraphics[width=1\linewidth]{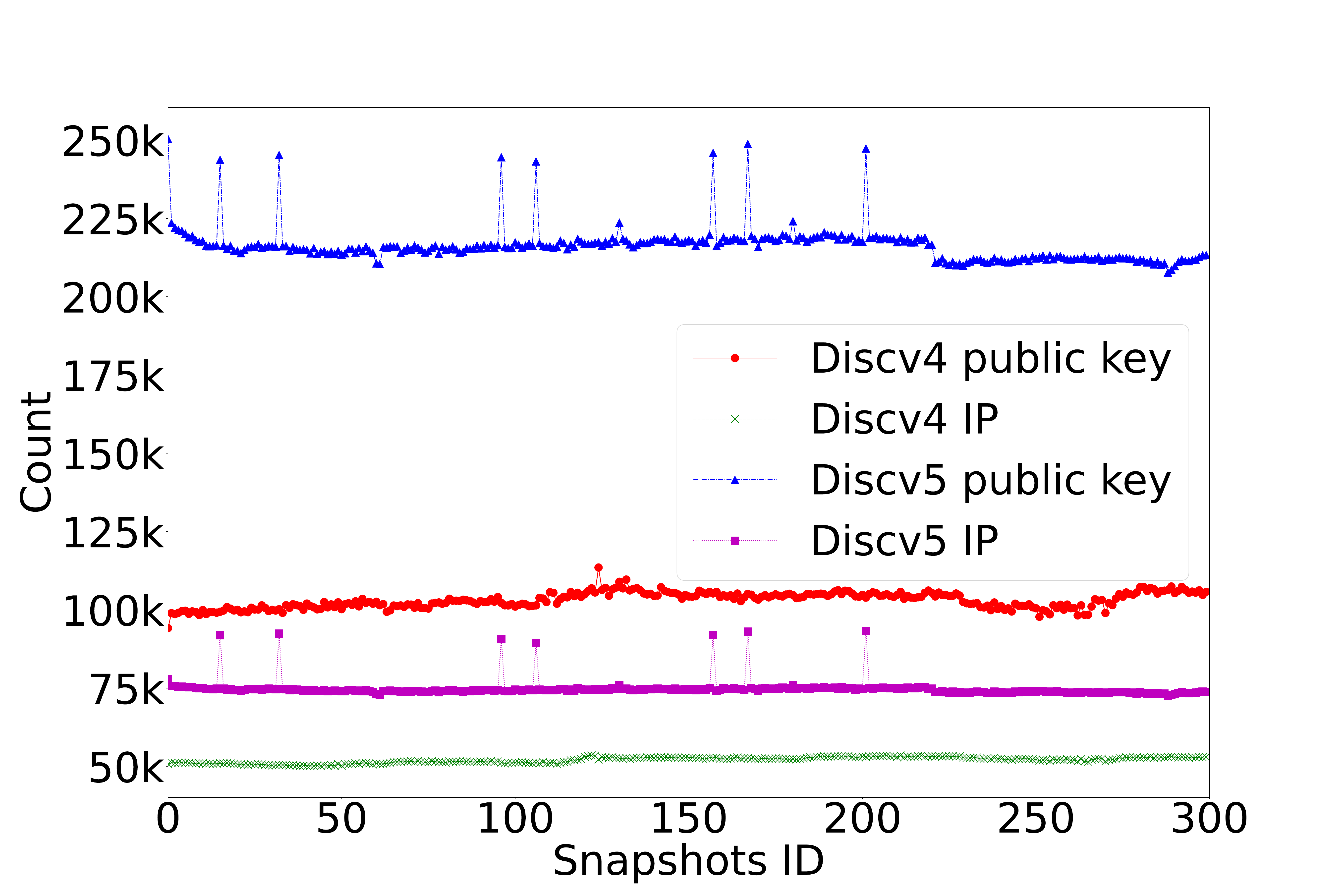}
  \caption{Line Chart of the Number of Public Keys and IPs from Snapshots after Continuous Measurement of Discv4 and Discv5 Networks.} 
  \label{fig:netmeasurement}
\end{figure}

As shown in Figure~\ref{fig:netmeasurement}, the daily average number of active nodes in Discv4, when nodes are identified by their public keys, generally remains around 100,000 while when identified by their IP addresses, the number typically stays around 50,000. 
In contrast, the daily average number of active nodes in Discv5, when identified by their public keys, is approximately 210,000 while when identified by their IP addresses, the number remains around 75,000. 
Whether in the Discv4 network or the Discv5 network, due to the scarcity of IP addresses, a large number of nodes are behind NAT. 
This is why the IP-based count is much smaller than the public key-based count. 
We cannot use the same IP address to identify users.

\begin{figure}[h]
  \centering
  \includegraphics[width=1\linewidth]{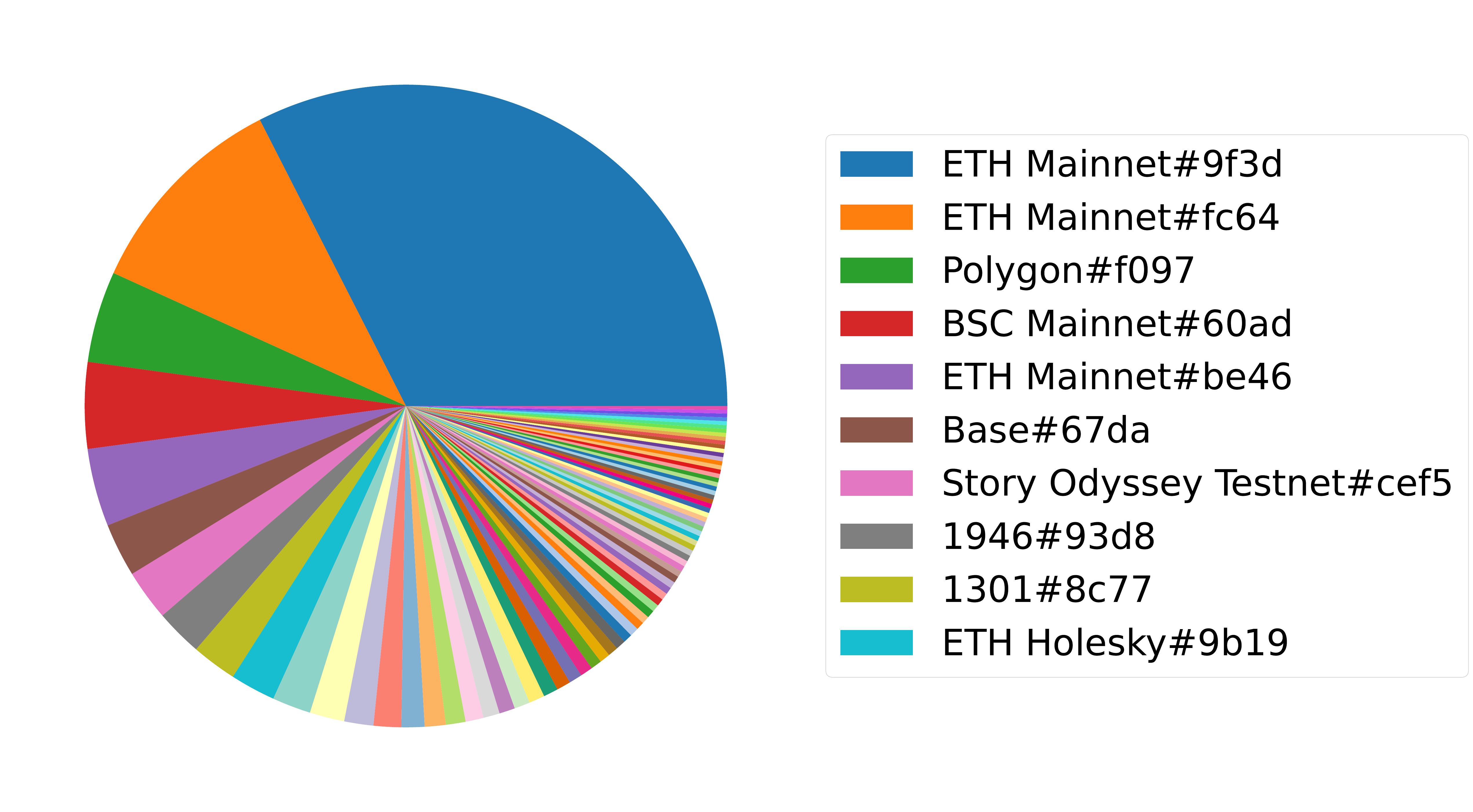}
  \caption{The pie chart below illustrates the proportion of node services in the Discv4 network over the measurement period. The larger the area of a segment in the pie chart, the greater the number of nodes providing that service. The number of nodes is identified using their public key as a unique identifier. The part before the hashtag represents the name of the service or network ID, while the part after the hashtag indicates the fork ID.}
  \label{fig:V4CountMap}
\end{figure}

The 300 snapshots of the Discv4 network cumulatively contain 30,923,495 duplicate nodes. 
Among these, we completed handshakes with 218,210 nodes and obtained their service information, as shown in Figure~\ref{fig:V4CountMap}. 
It can be observed that the services in the Discv4 network are highly diverse. 
Among them, the top five services—ETH Mainnet\#9f3d, ETH Mainnet\#fc64, Polygon\#f097, BSC Mainnet\#60ad, and ETH Mainnet\#be64—account for more than 50\% of the total.
Nodes that fail to complete the handshake and thus cannot obtain service information may be unable to do so due to reaching the peer limit, the node being behind NAT, issues in the TCP/IP network, or the use of alternative handshake methods.

\begin{figure}[h]
  \centering
  \includegraphics[width=1\linewidth]{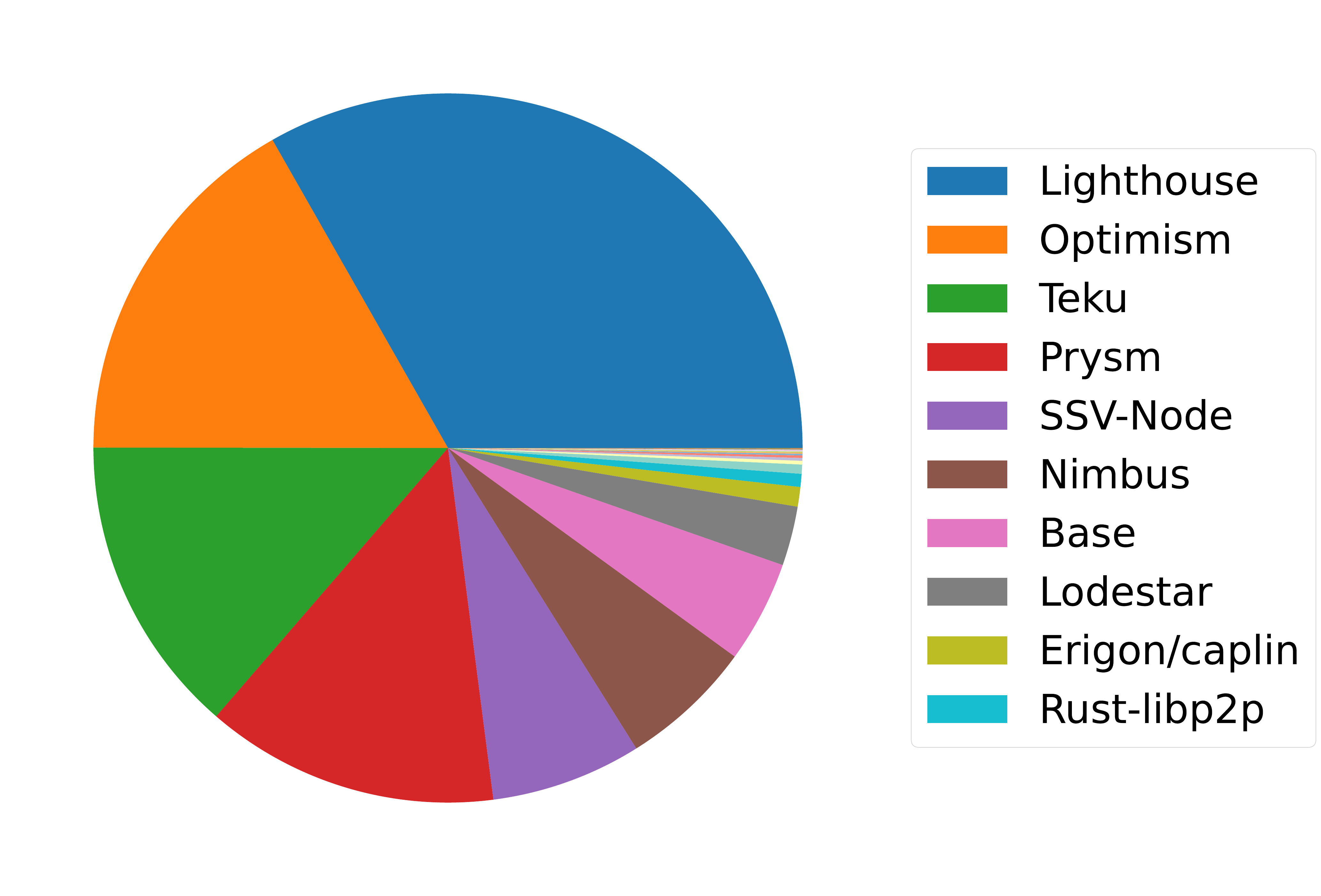}
  \caption{The proportion of Agents in the Discv5 network is distinguished using Agents due to the significant differences in the service handshake process within the Discv5 network. The larger the area of the pie chart, the higher the proportion of that Agent's count.}
  \label{fig:V5CountMap}
\end{figure}

A total of 64,894,147 potentially duplicate nodes are found across 300 Discv5 snapshots, of which 3,663,969 contain Agent information, as shown in Figure~\ref{fig:V5CountMap}. 
Compared to Discv4, the diversity of services in Discv5 is more pronounced due to the significant differences in the handshake process, and the degree of code reuse is lower than that of Discv4. 
Among the Agents with the largest proportions, Lighthouse, Teku, and Prysm are all Ethereum consensus service Agents. 
Optimism, on the other hand, is an Ethereum Layer 2 service Agent. 
The remaining nodes, for which Agent information could not be obtained due to protocol limitations, also cannot provide specific handshake details. 
This issue of missing information is similarly unresolved by other related works~\cite{Maeng2020,Maeng2021,Kim2018,Lucianna2021,Lee2020,Chen2020,Yue2019,Maeng2021Visualization,Zhenzhen2020,Mikel2024}.

\subsection{User Identity Integration Analysis}
\label{RQ3}
The Integrator integrates identity information from the data of the Crawler and Handshaker. 
From 300 snapshots, we discovered that 83 users used their public keys across a total of 485 different service nodes.

\begin{figure}[h]
  \centering
  \begin{minipage}[t]{0.48\linewidth}
    \centering
    \includegraphics[width=1\linewidth]{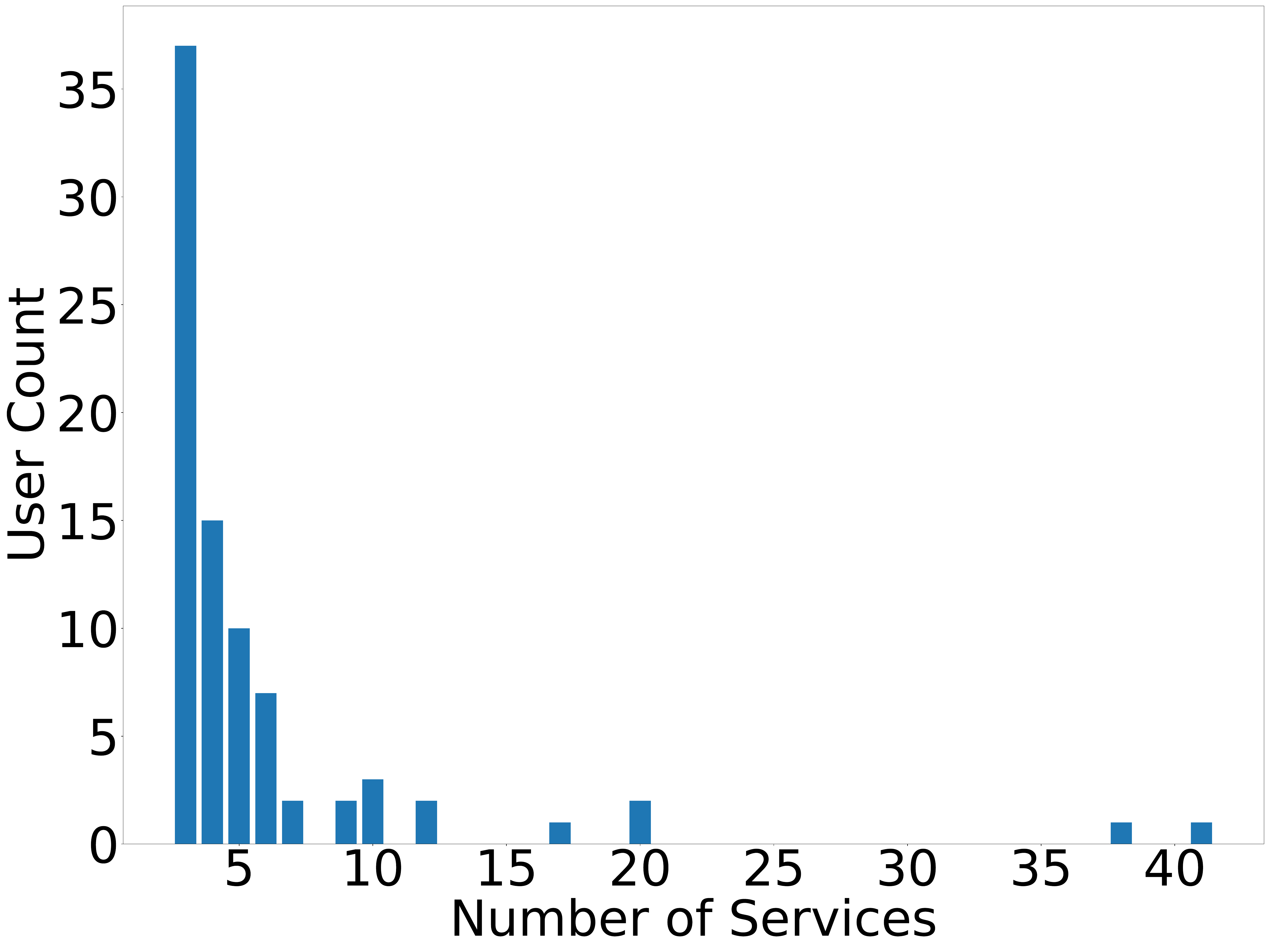}
    \caption{The distribution diagram of the number of services held by users, where the horizontal axis represents the number of services held by users, and the vertical axis indicates how many users hold that number of services.} 
    \label{fig:userservicedis}
  \end{minipage}
  \hfill
  \begin{minipage}[t]{0.48\linewidth}
  \centering
  \includegraphics[width=1\linewidth]{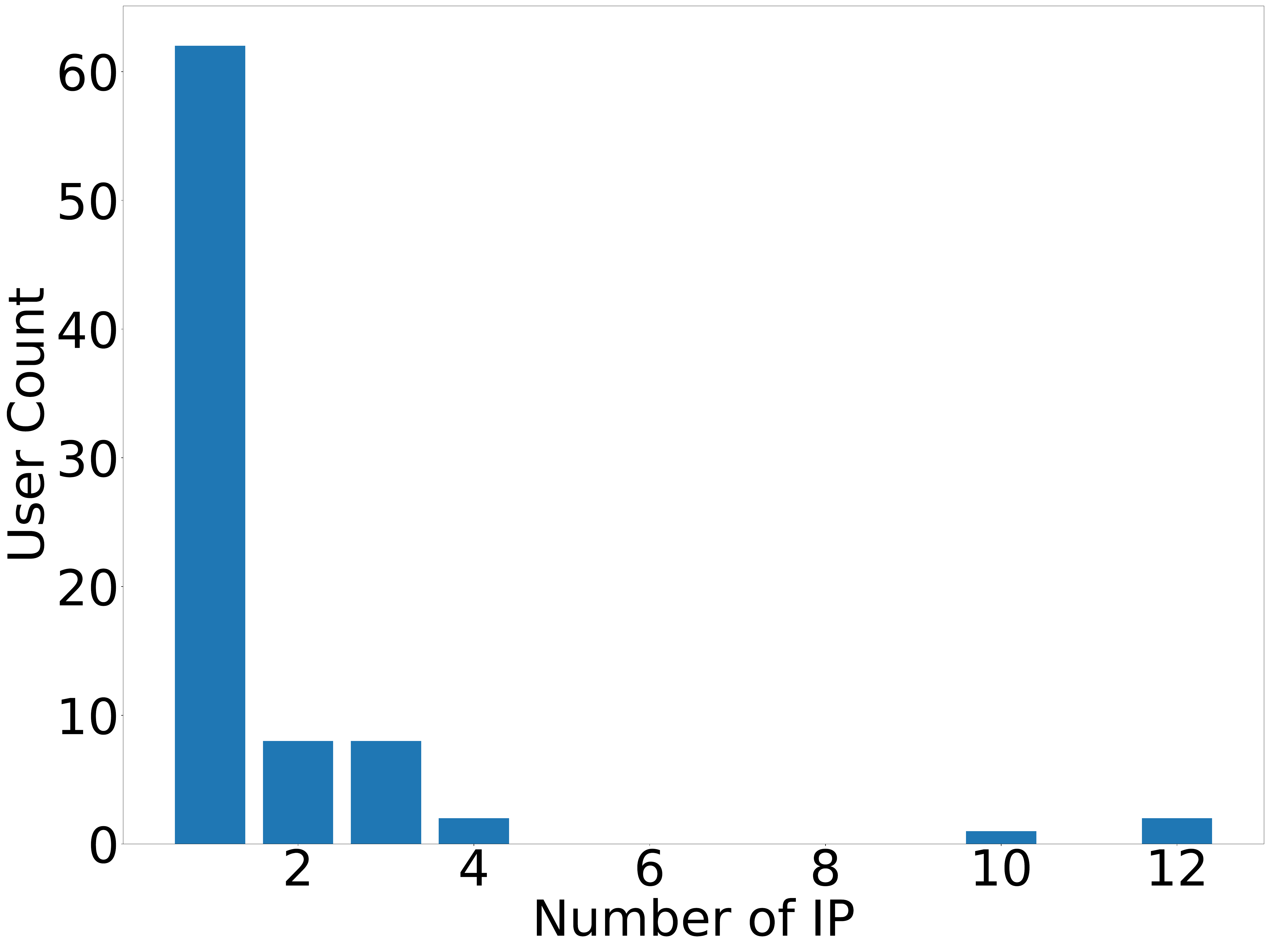}
  \caption{The distribution diagram of the number of IPs held by users, where the horizontal axis represents the number of IPs held, and the vertical axis indicates the number of users holding that number of IPs.}
  \label{fig:useripdis}
  \end{minipage}
\end{figure}

These 83 users reused their public keys, thereby compromising the network’s protection of their privacy, which allowed us to accurately identify them within the network.
As shown in Figure~\ref{fig:userservicedis}, approximately 74\% of users hold between 3 to 5 services. 
Among the remaining users, a smaller number hold a significantly higher number of services, with two users exhibiting extreme cases, holding 39 and 41 services, respectively.
As shown in Figure~\ref{fig:useripdis}, we found that 62 users with multiple services each used only a single IP address. 
Since we know we can handshake with these nodes, it can be inferred that the user either holds this IP address, or the NAT on the IP is bidirectional and bound to the user. 
The remaining 21 users held two or more IP addresses, indicating that these users have a larger number of IP resources. 
Specifically, one user owns 10 IP addresses, while two users each hold 12 IP addresses.

\begin{figure}[h]
  \centering
  \begin{minipage}[t]{0.48\linewidth}
    \centering
    \includegraphics[width=\linewidth]{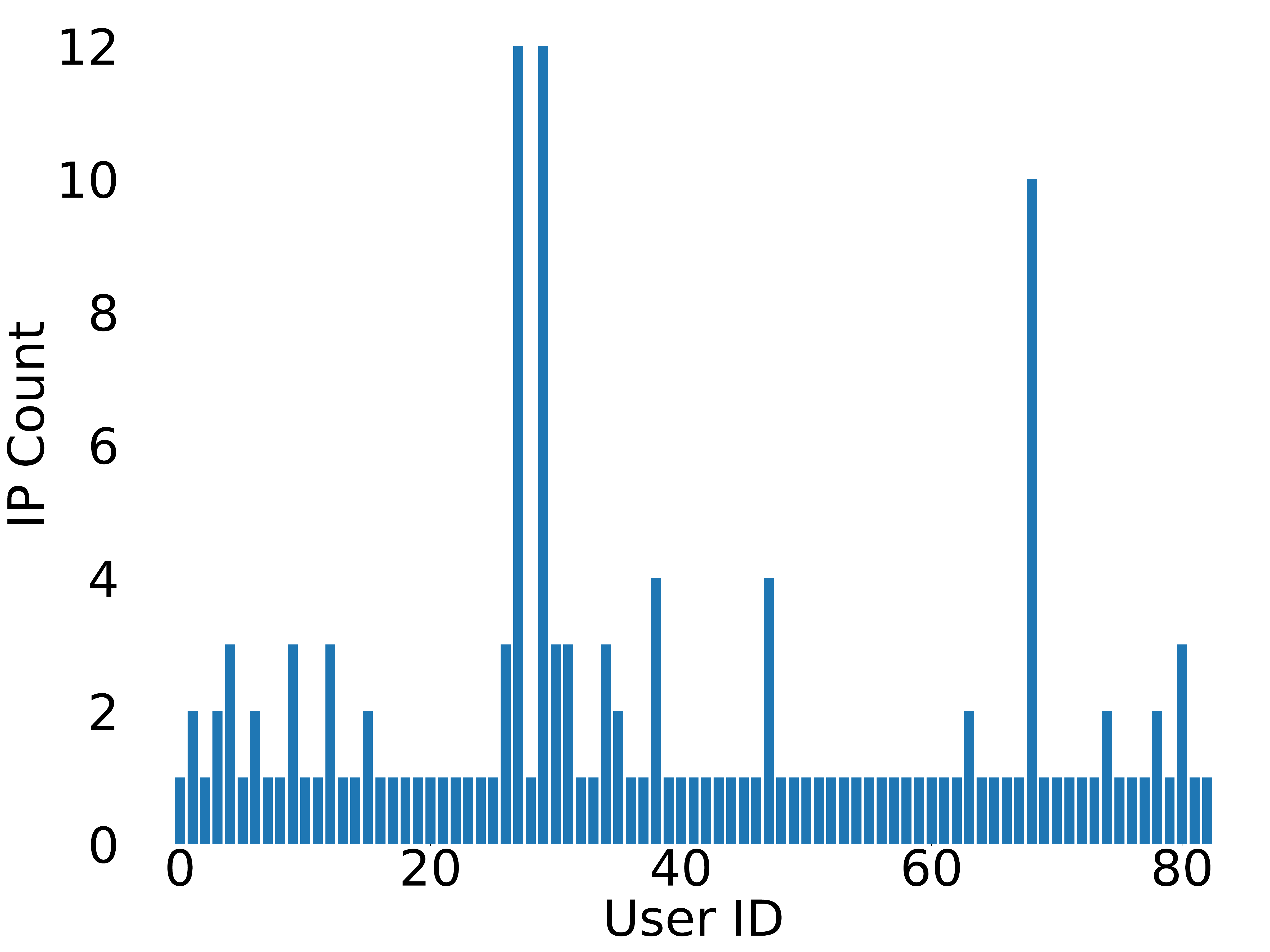}
    \caption{The distribution chart of the number of IPs held by users, where the x-axis represents user IDs and the y-axis denotes the number of IPs held by each user.} 
    \label{fig:useripcount}
  \end{minipage}
  \hfill
  \begin{minipage}[t]{0.48\linewidth}
    \centering
    \includegraphics[width=\linewidth]{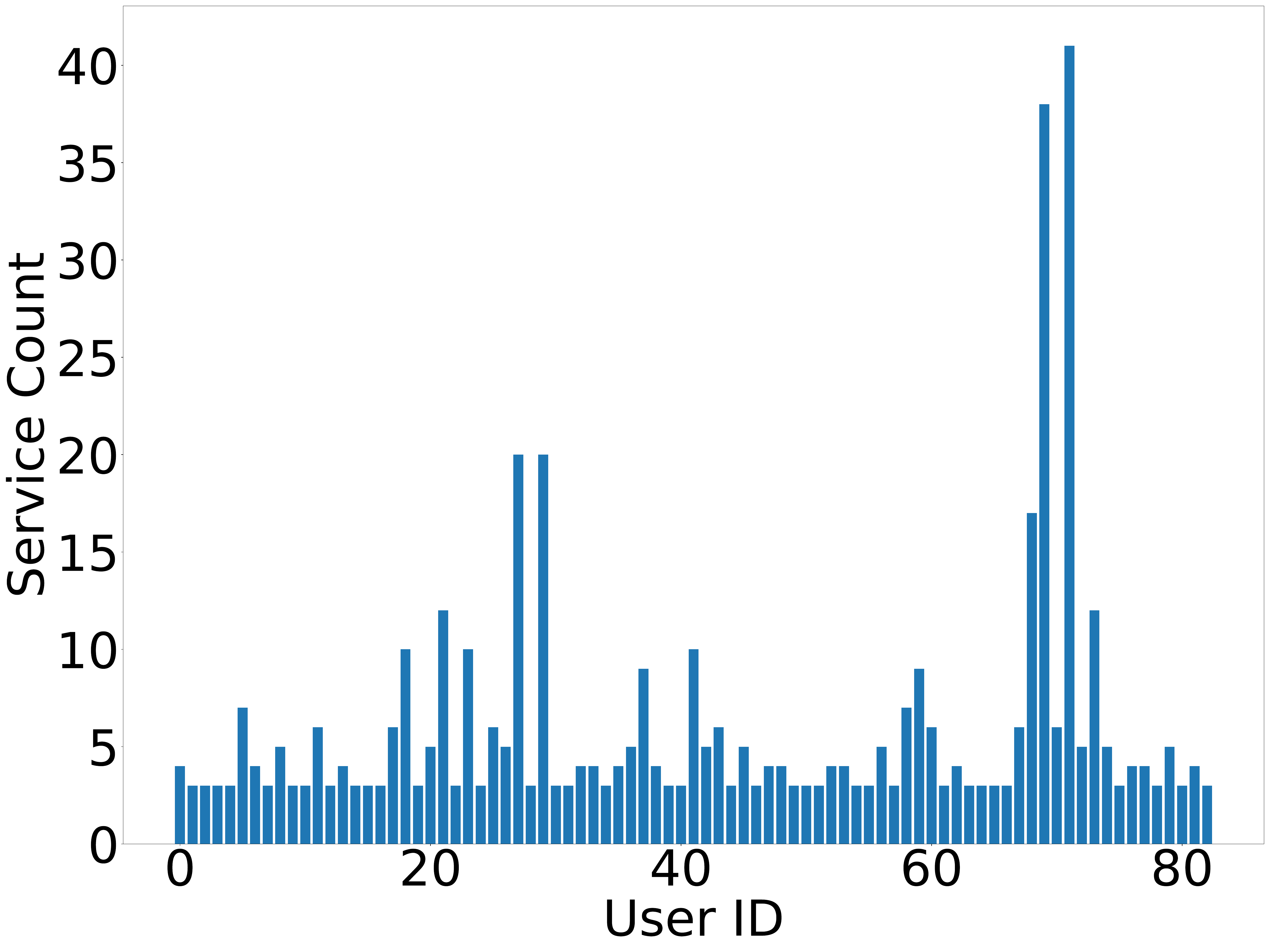}
    \caption{The distribution chart of the number of Services held by users, where the x-axis represents user IDs and the y-axis indicates the number of Services associated with each user.} 
    \label{fig:userservicecount}
  \end{minipage}
\end{figure}

By analyzing user behavior based on user IDs, as illustrated in Figure~\ref{fig:useripcount} and Figure~\ref{fig:userservicecount}, it can be observed that the majority of users hold a relatively small number of IPs and Services. 
Only a minority of users possess multiple IPs and Services. 
Most users tend to deploy multiple distinct Services on the same IP address, whereas a smaller group of users choose to acquire multiple IPs to distribute their Services across different addresses.

\begin{figure}[h]
  \centering
  \includegraphics[width=1\linewidth]{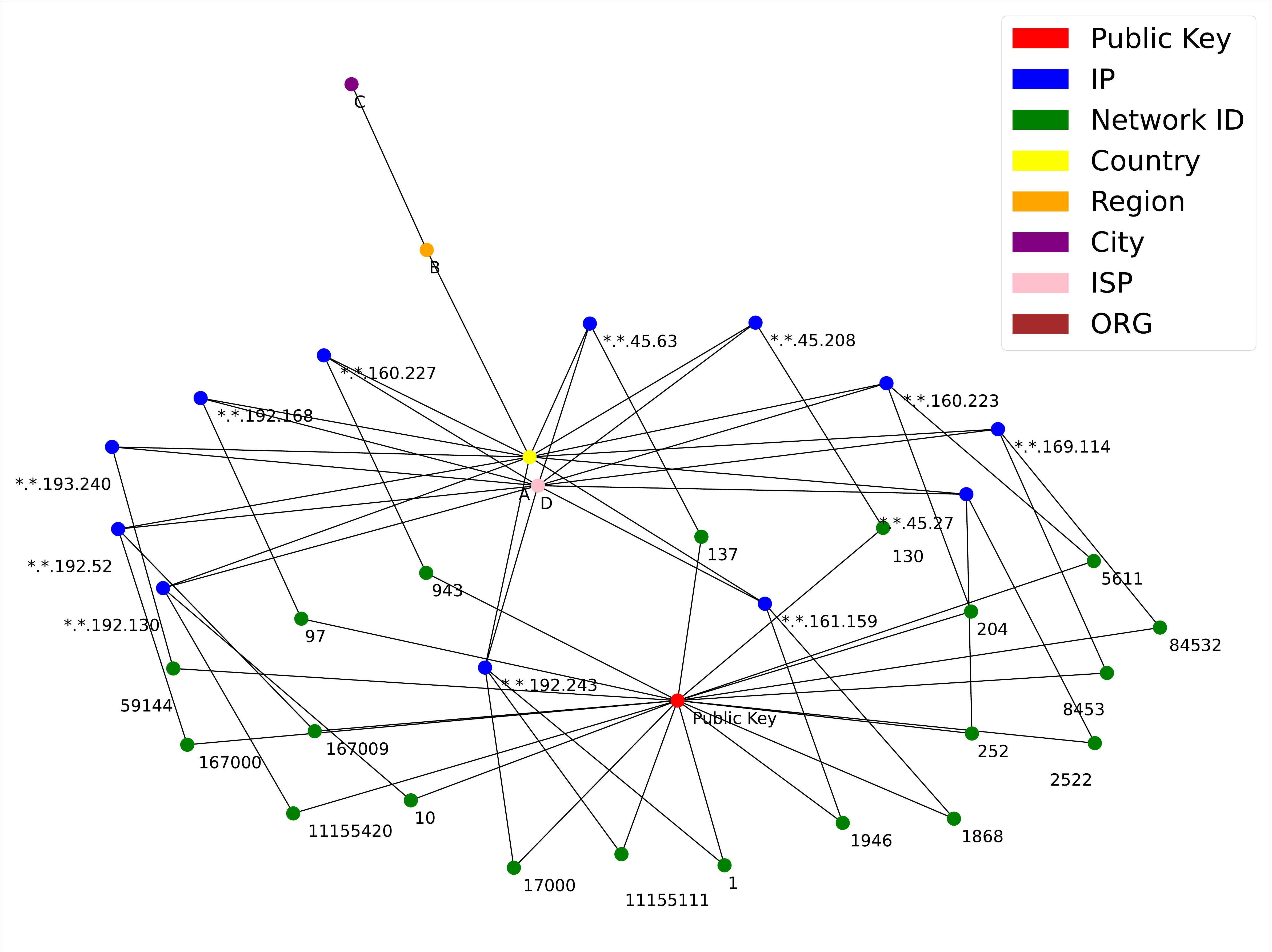}
  \caption{The graph illustrating the IPs, Services, and public network information associated with User27, as reconstructed by the Graph-Based Identity Association Algorithm.} 
  \label{fig:user27}
\end{figure}

After analyzing the IP usage data of these users from public sources, we find that a portion of the users directly utilize cloud service providers such as Google Cloud, Hetzner, AWS, and Latitude.sh. 
Another group of users relies on broadband service providers like Jupiter Telecommunication Co. Ltd and Chubu Telecommunications Co., Inc. 
Some users are clients of companies that provide node instance servers specifically for blockchain users. 
Given that the public key has no risk of forgery or collision, if these users are involved in activities that violate local laws and regulations, it is possible to retrieve their personal information from their network service providers. 
As a result, these users do not fully benefit from the privacy offered by the blockchain network.
For example, as shown in Figure~\ref{fig:user27}, we apply the Graph-Based Identity Association Algorithm to reconstruct the user information of User27. 
The analysis reveals that the user is located in Country A, Region B, and City C, and leverages network services provided by Company D to manage their owned Services.

\section{Discussion And Limitations}
\label{sec:disLimit}
This section presents the experimental results analysis and  limitations.

The experimental results indicate that users fail to adhere to cryptographic design specifications. 
While the network protocol itself is secure, users’ improper cryptographic practices create significant security and privacy risks. 
Unlike studies~\cite{Maeng2020,Maeng2021,Kim2018,Lucianna2021,Lee2020,Chen2020,Yue2019,Maeng2021Visualization,Zhenzhen2020,Mikel2024,DISCNG} focused on interpreting results, our experiments emphasize analyzing the root causes and consequences of these issues. 
Unlike research~\cite{Eisenbarth2022,XU2020,Marcus2018,Sebastian2019,Hwanjo2023,TaotaoWang2021Ethna,Kai2021,Zhao2024DEthna,Dominic2025,Caspar2021,Mikel2024,DISCNG,LSDAttack} that examines the security of protocols and mechanisms, our study highlights the insecurity arising from user behavior, as indicated by the data. 
The experiments are reproducible, and the findings can be applied to similar blockchain or distributed networks.

This study has three limitations, though they have minor impacts. 
First, the snapshots from the crawler represent a subset of the network nodes during a given period, as new nodes continuously join and leave the distributed network. 
Thus, capturing the entire network in a short time is impractical, a challenge faced by other studies. 
Second, the Handshaker cannot collect handshake information for all services. 
Many nodes are visible at the TCP/IP layer, but service data is inaccessible due to differences in handshake processes, a common issue in other studies that focus on well-known services. 
Third, limitations imposed by public databases and telecom providers prevent access to users’ personal identity information. 
Thus, only telecom providers are identifiable, but detailed personal data is unavailable.

\section{Conclusion}
\label{sec:conclusion}
This paper makes three fundamental contributions to advancing decentralized network security and blockchain privacy protection:
First, we pioneer the EGNInfoLeaker framework – the first specialized system for detecting public key reuse vulnerabilities in Discv4/Discv5 protocols. Our tool establishes new capabilities for identifying this critical but previously overlooked attack vector in mainstream decentralized networks.
Second, through large-scale behavioral forensics, we demonstrate how malicious actors exploit public key reuse to bypass network anonymity safeguards. Our analysis proves these practices enable identity correlation attacks that simultaneously compromise both node privacy and network security architectures.
Third, we develop a novel de-anonymization profiling algorithm that reconstructs real-world identities from protocol metadata. This methodology not only exposes concrete privacy risks but also provides a benchmark for evaluating anonymity in decentralized systems.
These breakthroughs collectively reshape understanding of privacy-security tradeoffs in blockchain networks. By exposing the public key reuse threat model and delivering the first operational detection system, our work enables next-generation protections for decentralized protocols. The empirical evidence and analytical tools presented create new pathways for developing accountable anonymity in peer-to-peer networks while preserving core Web3 values.

\bibliographystyle{plain}
\bibliography{myref.bib}
\end{document}